# Improved Superconducting Properties in Nanocrystalline Bulk $MgB_2$

A. Gümbel, J. Eckert*, G. Fuchs*, K. Nenkov, K.-H. Müller, and L. Schultz

*IFW Dresden, Institute of Metallic Materials, Helmholtzstr. 20, D-01069 Dresden, Germany*

## Abstract

Highly dense nanocrystalline $MgB_2$ bulk superconductors with distinctly improved pinning were prepared by mechanical alloying of Mg and B powders and hot compaction at ambient temperatures. The nanocrystalline samples reveal high $j_c = 10^5$ A/cm$^2$ at 20 K and 1 T together with a strongly shifted irreversibility line towards higher fields resulting in $H_{irr}(T) \sim 0.8\ H_{c2}(T)$, whereas typically $H_{irr}(T) \sim 0.5\ H_{c2}(T)$ is observed for bulk untextured samples. These values exceed that of all other so far reported bulk samples and are in the range of the values of thin films. The improved pinning of this material which mainly consists of spherical grains of about 40-100 nm in size is attributed to the large number of grain boundaries.

* *Correspondence and requests for materials should be addressed to J.E. (e-mail: j.eckert@ifw-dresden.de) or G.F. (e-mail: fuchs@ifw-dresden.de).*

The recent discovery of superconductivity in $MgB_2$[1] was completely unexpected as the preparation and the structure of this material have already been described almost 50 years ago[2]. In contrast to high $T_c$ superconductors, grain boundaries in $MgB_2$ are not acting as impediments for superconducting currents[3,4]. The combination with high critical fields $H_{c2}$ and irreversibility fields $H_{irr}$ of more than 30 T and a critical current density $j_c$ of 3MA/cm$^2$ at 4.2 K and 1 T which are found for thin films[5,6] render $MgB_2$ promising for technical applications. However, in commercial powders and conventional sintered polycrystalline bulk $MgB_2$ samples and wires the above mentioned properties are reduced due to a worse pinning behaviour. Here, we describe a preparation route for nanocrystalline bulk samples with distinctly improved pinning via mechanical alloying of Mg and B powders and hot compaction. The samples reveal high $j_c = 10^5$ A/cm$^2$ at 20 K and 1 T together with high $H_{c2}$ and $H_{irr}$ values which exceed that of all other so far reported bulk samples and are in the range of the values of thin films.

Mechanical alloying is a well established method to prepare metastable amorphous, quasicrystalline or nanocrystalline materials[7]. Starting materials and milling balls are placed together in a milling vial, which is exposed to a strong vibrational or rotational acceleration causing impacts of milling balls leading to repeated fractioning and cold welding of powder particles. Depending on the initial materials the milling conditions can be adjusted so that the desired phase forms via solid-state reaction at moderate temperatures[7].

In this study Mg (99.8 %) and amorphous B (99.9+ %) powders with mass-ratio according to $MgB_2$ were filled under purified Ar-atmosphere into a tungsten carbide (WC) vial containing WC balls. The milling was performed on a planetary ball mill for different times $t_m$ = 20–100 hours using a ball-to-powder mass ratio of 36. While for short milling times the precursor powder consists of a distinct $MgB_2$ phase fraction together with unreacted nanocrystalline Mg and amorphous B, it is also possible to achieve complete $MgB_2$ formation after 100 hours of milling.

Batches of about 0.15 g of precursor powders were compacted in a hot press at 700°C and 640 MPa for pressing times $t_p$ = 10 – 90 minutes leading to pellets with a diameter of 10 mm and a height of 0.8 mm. The density reached more than 80 % of the theoretical value after compaction for 10 min and about 95 % after compaction for 60 min or longer. For resistivity and magnetic measurements the pellets were cut into stripes of 9 mm length and 1.5 mm width. In this work a sample with $T_c$ = 34.5 K, milled for 20 hours and pressed at 700°C and 640 MPa for 10 min is described.

Fig. 1 shows X-ray diffraction patterns of as-milled powder and the compacted sample revealing the phase formation of $MgB_2$. Obviously during milling a certain amount of $MgB_2$ already forms. Complete reaction is achieved by subsequent annealing inducing a strong exothermic reaction starting well below 500 °C which is completed at 650 °C as proven by differential scanning calorimetry (inset in Fig. 1). To ensure this phase formation during hot compaction, similar annealing conditions were used. No remainders of the starting materials and no indication of texture could be detected in the compact which was investigated by powder pattern structure refinement. Beside a small amount of 3 vol. % MgO there is only little contamination of 0.3 vol. % WC stemming from the milling tools visible, yielding a $MgB_2$ phase fraction of more than 96 vol. %. The broad diffraction peaks indicate small dimensions of coherent scattering domains of about 15 nm. This can be regarded as a minimum bound for the grain size[8].

Fig. 2 shows typical scanning electron microscope images of the compacted sample with a density of approximately 2.1 g/cm$^3$ which is about 80 % of the theoretical value[2]. The material mainly consists of spherical grains of about 40-100 nm in size, larger than the domain size evaluated from x-ray analysis. This indicates a significant amount of strain and structural defects.

The field-dependent superconducting properties were probed by *ac* susceptibility in applied fields up to 9 T and resistance (standard four point method) measurements up to 17.5 T. The upper critical field $H_{c2}$ and the irreversibility field $H_{irr}$ were determined at 90 % of the normal-state resistance and zero resistance, respectively. The temperature dependence of $H_{c2}$ and of the irreversibility field $H_{irr}$ which corresponds at a given temperature to the highest field for non-zero critical currents are plotted in Fig. 3 together with data of high quality bulk samples[9] and thin films[5]. Compared to the bulk sample with micrometer grain size[9] the nanocrystalline sample reported here reveals much higher irreversibility fields especially at low temperatures, whereas the $H_{c2}$ data of these sample are comparable. The strong shift of the irreversibility line $H_{irr}(T)$ of the nanocrystalline sample towards higher magnetic fields resulting in a reduced separation between $H_{irr}(T)$ and $H_{c2}(T)$ is due to improved flux pinning and may be caused by the enhanced number of grain boundaries. This interpretation is supported by the comparison with data for a nanocrystalline thin film[5] shown in Fig. 3b for two field orientations. Despite its lower $T_c$ of about 31 K this c-axis oriented thin film which was alloyed with oxygen has strongly enhanced $H_{c2}(T)$ and $H_{irr}(T)$ values which surpass those of all bulk samples below about 27.5 K. One possible reason for the strong pinning found for thin films is their small grain size of about 10 nm which is considerably smaller than for our samples. It is worth to note that the irreversibility line and $H_{c2}(T)$ of our sample almost coincide with the corresponding data of the film in the unfavourable orientation with the applied field perpendicular to the film plane.

The strong disorder in the oxygen-alloyed thin film results in a very high resistivity of ρ(40K) = 360 μΩcm and a so-called dirty limit behaviour of $H_{c2}(T)$ with a linear $H_{c2}(T)$ dependence near $T_c$. Typical bulk $MgB_2$ samples have a much smaller resistivity of ρ(40K) ~ 5 μΩcm (tending downwards even to ρ(40K) ~ 0.4 μΩcm for wires[10,11]) and a positive curvature of $H_{c2}(T)$ near $T_c$ which is characteristic for an effective two-band superconductor[12] in the clean limit. The value of ρ(40K) = 46 μΩcm and the slight positive curvature of $H_{c2}(T)$ near $T_c$ of the investigated nanocrystalline samples suggest superconductivity near the transition from clean to dirty limit.

The critical current density $j_c$ shown in Fig. 4 was determined from *dc* magnetization loops at temperatures $T$ = 20–30 K using a SQUID magnetometer. The standard Bean model[13] was applied to evaluate the field dependence of $j_c$ at different temperatures from the hysteresis of the magnetization and the sample size. At temperatures below 26 K flux jumps were observed at low applied fields resulting from thermomagnetic instabilities. Therefore, $j_c$ could be determined only in a limited range up to about $10^5$ A/cm$^2$. Nevertheless, it is evident from Fig. 4 that a critical current density of $j_c = 1{\cdot}10^5$ A/cm$^2$, which is a standard requirement for technical applications, still persists at an applied field of 2.1 T at 20 K which is even higher than found for thin films[5].

The present work reveals that it is possible to prepare highly dense nanocrystalline $MgB_2$ bulk material with excellent values of $H_{irr}(T)$ and $H_{c2}(T)$. $T_c$ was found to be lowered by several K which may be ascribed to possible contamination effects from the milling tools or increased internal strain. Compared to bulk material prepared by the standard technique, the irreversibility line of this nanocrystalline material is strongly shifted towards higher fields resulting in $H_{irr}(T) \sim 0.8\ H_{c2}(T)$, whereas typically $H_{irr}(T) \sim 0.5\ H_{c2}(T)$ is observed for bulk untextured samples. Furthermore, a high critical current density of $j_c = 1 \cdot 10^5$ A/cm$^2$ was found at 20 K in fields up to 2 T. The improved pinning of this material is most probably due to a large number of grain boundaries as already predicted previously[14,15]. Hence, further studies on mechanically alloyed nanocrystalline $MgB_2$ are promising to develop material suitable for future applications.

**Acknowledgements**

We thank H. Schulze for technical guidance, and B. de Boer, I. Mönch and B. Zhao for the realization of the SEM investigations.

## Figure captions

**Fig. 1.** X-ray diffraction patterns of the as-milled powder and the hot compacted sample. Peaks of $MgB_2$ are indexed, peaks of MgO or WC are marked by squares and circles, respectively. In the as-milled powder (lower pattern) mainly peaks of nanocrystalline Mg are visible but also distinct peaks of $MgB_2$ arise. The compacted sample (upper pattern) is almost single phase $MgB_2$ with a minor MgO fraction of 3 vol.%. In both patterns the WC wear debris from the milling tools appears which is 0.3 vol.% in the compacted sample. Inset, Differential scanning calorimetry (DSC) scan at a rate of 20 K/min of the as-milled powder revealing the transformation of the precurser powder into $MgB_2$.

**Fig. 2.** Typical scanning electron micrographs of a milled and subsequently hot pressed sample. The high resolution reveals a very fine microstructure consisting of grains with a size of 40–100 nm and nearly uniform spherical shape.

**Fig. 3.** Temperature dependence of the upper critical field $H_{c2}$ and of the irreversibility field $H_{irr}$ of $MgB_2$ samples. **a**, Comparison of $H_{c2}$ (full circles) and $H_{irr}$ (open circles) of the mechanically alloyed sample with data for a bulk sample[9]. $H_{c2}$ and $H_{irr}$ of the mechanically alloyed sample were determined at 90 % of the normal state resistivity and at zero resistivity, respectively. $H_{irr}$ data derived from the peak value of the imaginary part of the *ac* susceptibility are marked by triangles and are comparable with the resistively determined data. $H_{c2}$ and $H_{irr}$ of the bulk sample are shown as full and dotted lines,

respectively, without symbols. **b**, Comparison of $H_{c2}$ (full circles) and $H_{irr}$ (open circles) of the same mechanically alloyed sample with data for a thin film[5] for two field orientations (H parallel and perpendicular to the film plane referred to as H || and H ⊥, respectively). $H_{c2}$ and $H_{irr}$ of the thin film are shown as full and dashed lines, respectively, without symbols.

**Fig. 4.** Critical current density $j_c$ at different temperatures T = 20, 24, 26, 28 and 30 K. For the calculation the standard Bean model for a plate in a perpendicular field $j_c(H) = 20 \cdot \Delta M(b-b^2/3 \cdot l)$ was used, where ΔM is the difference of magnetization (in emu/cm$^2$) measured for ascending and descending applied field, $b$ and $l$ are the sample width and length in cm, respectively. This model implies that the superconductor is in the critical state, i.e. the field gradient within the superconductor corresponds to its critical current density. At temperatures T < 26 K and low magnetic fields the critical state can be destroyed by flux jumps resulting from thermomagnetic instabilities. Inset: Magnetization loop at 24 K (SQUID data) showing two flux jumps at low applied fields.

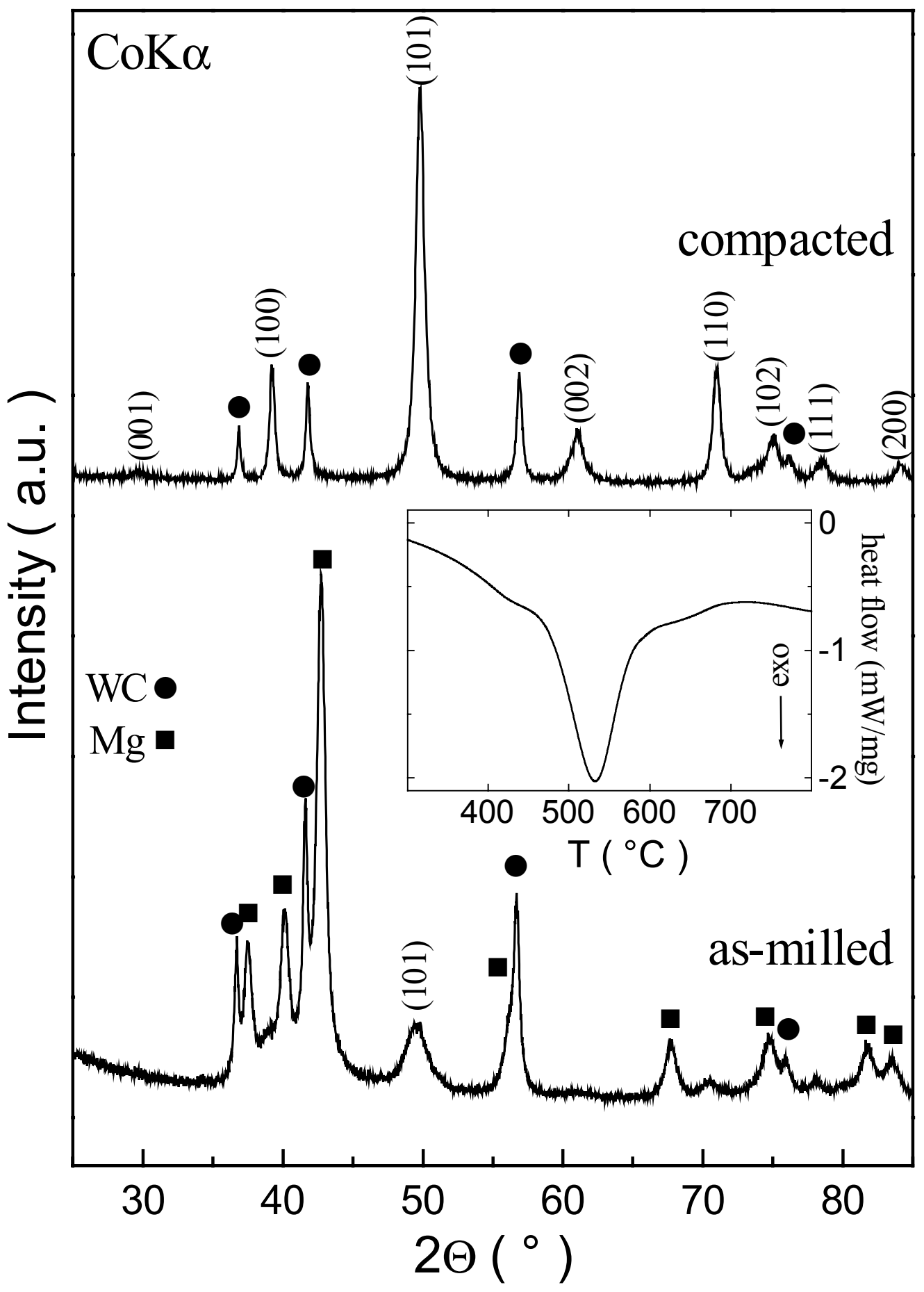

Figure 1 Gümbel et al.

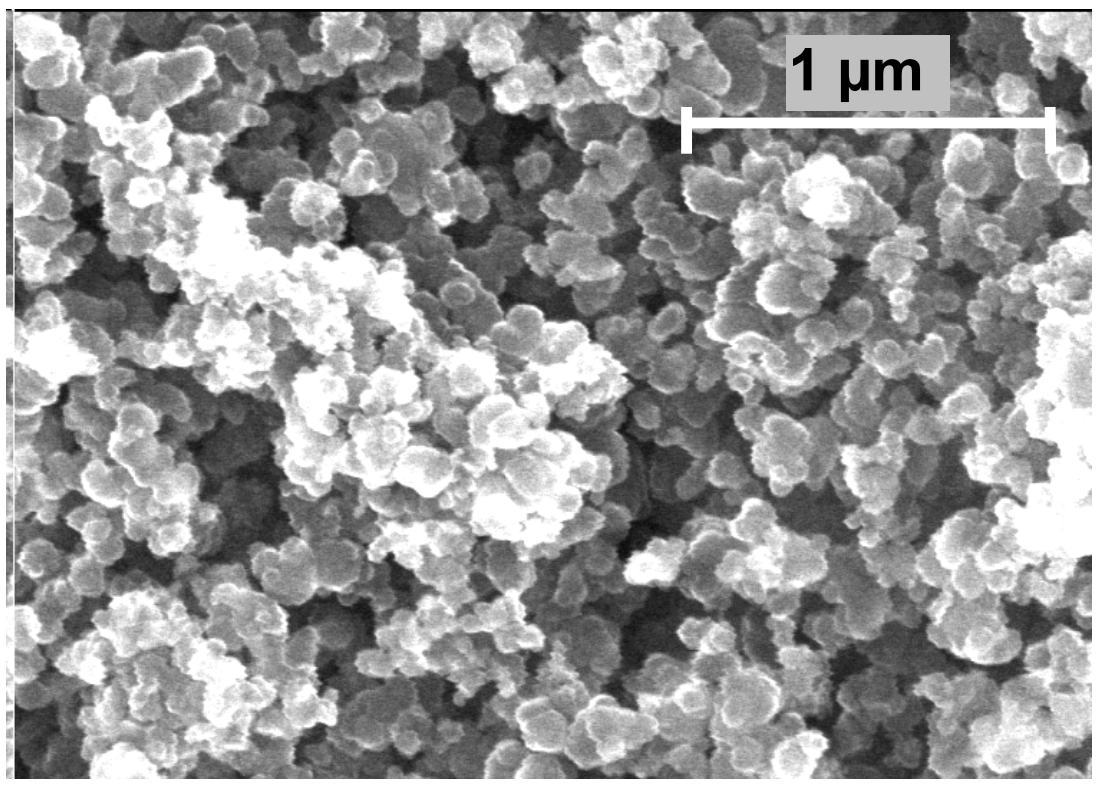

Figure 2 Gümbel et al.

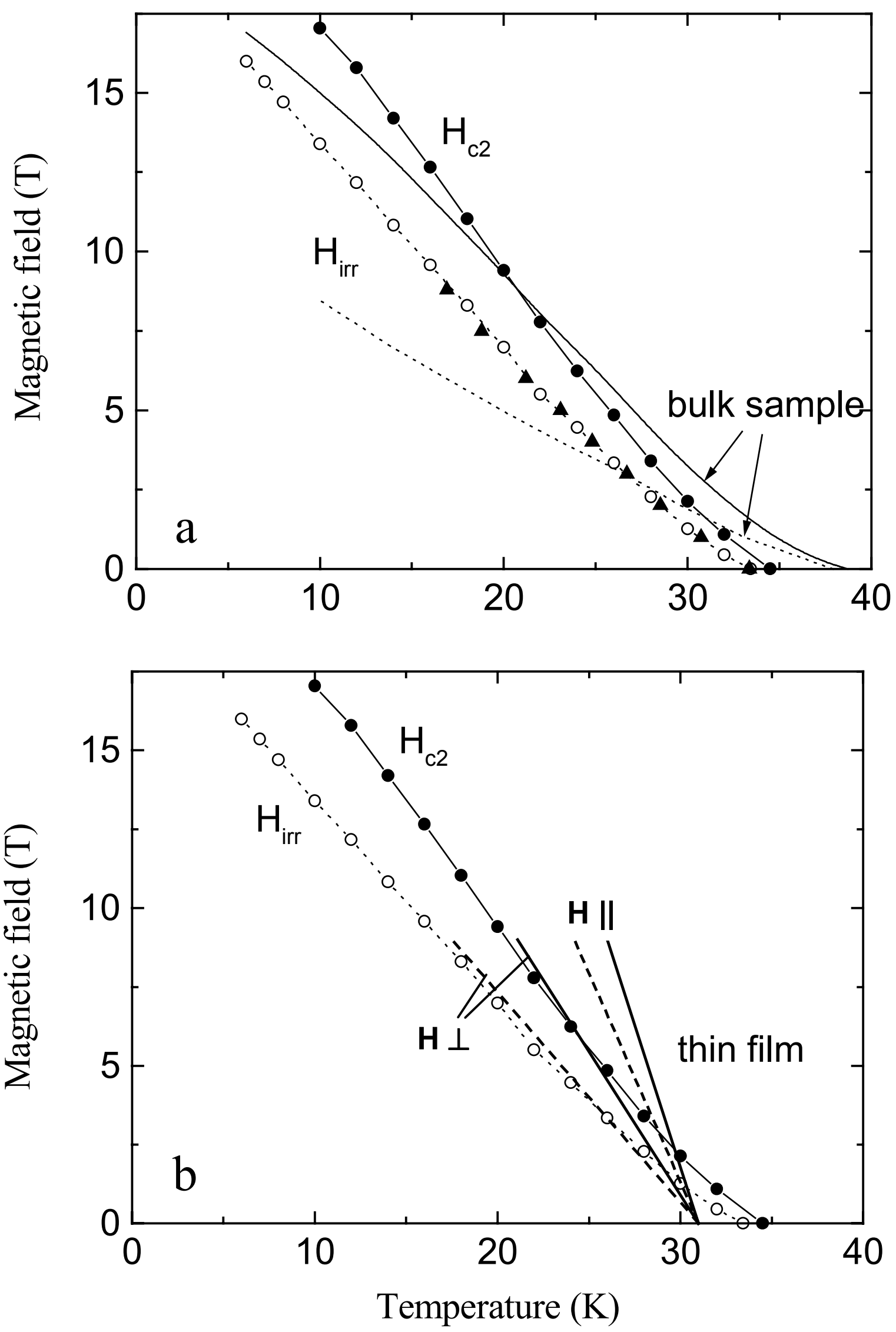

Figure 3 Gümbel et al.

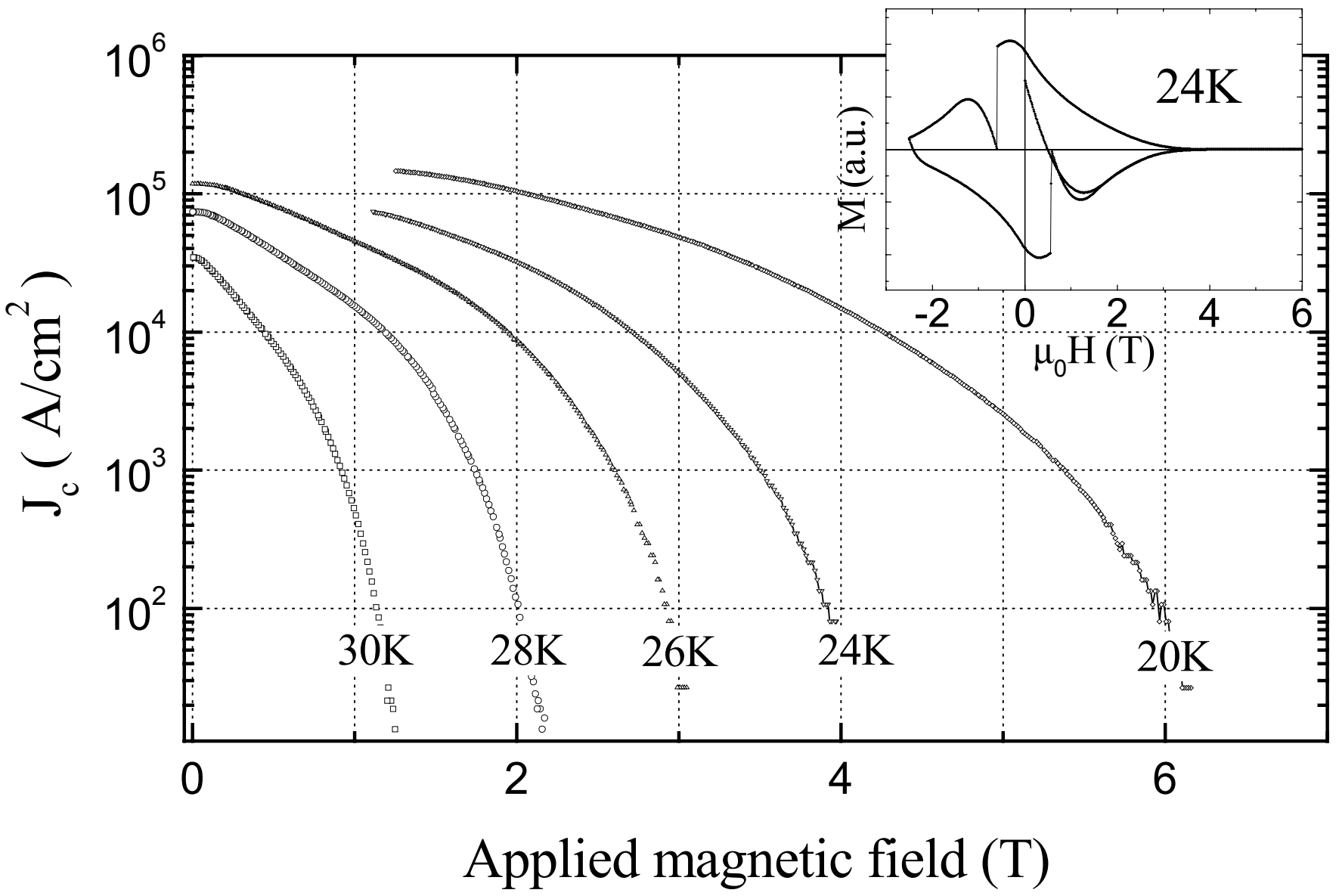

Figure 4 Gümbel et al.